\documentstyle[12pt]{article} 
\input psfig.sty
\def\Re{{\cal R \mskip-4mu \lower.1ex \hbox{\it e}\,}}
\def\Im{{\cal I \mskip-5mu \lower.1ex \hbox{\it m}\,}}
\def\nn{\noindent}
\def\ie{{\it i.e.}}
\def\eg{{\it e.g.}}

\def\etal{{\it et al.}}

\def\sub#1{_{\lower.25ex\hbox{$\scriptstyle#1$}}}
\def\sul#1{_{\kern-.1em#1}}
\def\sll#1{_{\kern-.2em#1}}  
\def\sbl#1{_{\kern-.1em\lower.25ex\hbox{$\scriptstyle#1$}}}
\def\ssb#1{_{\lower.25ex\hbox{$\scriptscriptstyle#1$}}}
\def\sbb#1{_{\lower.4ex\hbox{$\scriptstyle#1$}}}

\def\to{\rightarrow}
\def\mh{\ifmmode m\sbl H \else $m\sbl H$\fi}
\def\mch{\ifmmode m_{H^\pm} \else $m_{H^\pm}$\fi}
\def\mt{\ifmmode m_t\else $m_t$\fi}
\def\mc{\ifmmode m_c\else $m_c$\fi}
\def\mz{\ifmmode M_Z\else $M_Z$\fi}
\def\mw{\ifmmode M_W\else $M_W$\fi}
\def\mws{\ifmmode M_W^2 \else $M_W^2$\fi}
\def\mhs{\ifmmode m_H^2 \else $m_H^2$\fi}   
\def\mzs{\ifmmode M_Z^2 \else $M_Z^2$\fi}
\def\mts{\ifmmode m_t^2 \else $m_t^2$\fi}
\def\mcs{\ifmmode m_c^2 \else $m_c^2$\fi}
\def\mchs{\ifmmode m_{H^\pm}^2 \else $m_{H^\pm}^2$\fi}
\def\ztwo{\ifmmode Z_2\else $Z_2$\fi}
\def\zone{\ifmmode Z_1\else $Z_1$\fi}
\def\mtwo{\ifmmode M_2\else $M_2$\fi}
\def\mone{\ifmmode M_1\else $M_1$\fi}
\def\tb{\ifmmode \tan\beta \else $\tan\beta$\fi}
\def\xw{\ifmmode x\sub w\else $x\sub w$\fi}
\def\ch{\ifmmode H^\pm \else $H^\pm$\fi}
\def\lum{\ifmmode {\cal L}\else ${\cal L}$\fi}
\def\inpb{\ifmmode {\rm pb}^{-1}\else ${\rm pb}^{-1}$\fi}
\def\infb{\ifmmode {\rm fb}^{-1}\else ${\rm fb}^{-1}$\fi}
\def\epem{\ifmmode e^+e^-\else $e^+e^-$\fi}
\def\ppb{\ifmmode \bar pp\else $\bar pp$\fi}

\def\bsg{\ifmmode b\rightarrow s\gamma \else $b\rightarrow s\gamma$\fi}
 
\newskip\zatskip \zatskip=0pt plus0pt minus0pt
\def\matth{\mathsurround=0pt}

\def\atversim#1#2{\lower0.7ex\vbox{\baselineskip\zatskip\lineskip\zatskip
  \lineskiplimit 0pt\ialign{$\matth#1\hfil##\hfil$\crcr#2\crcr\sim\crcr}}}

\renewcommand{\thefootnote}{\fnsymbol{footnote}}

\begin{document} \begin{titlepage} 
\setcounter{page}{1}
\thispagestyle{empty}
\rightline{\vbox{\halign{&#\hfil\cr
&SLAC-PUB-7453\cr
&April 1997\cr}}}
\vspace{0.8in} 
\begin{center}

{\Large\bf Searching For Anomalous $\tau \nu W$ Couplings}
\footnote{Work supported by the Department of 
Energy, contract DE-AC03-76SF00515.} 
\medskip

\normalsize THOMAS G. RIZZO
\\ \smallskip
{\it {Stanford Linear Accelerator Center\\Stanford University, 
Stanford, CA 94309}}\\ 

\end{center}

\begin{abstract} 

The capability of current and future measurements at low and high energy 
$e^+e^-$ colliders to probe for the existence of anomalous, CP conserving, 
$\tau \nu W$ dipole moment-type couplings is examined. At present, constraints 
on the universality of the tau charged and neutral current interactions as well 
as the shape of the $\tau \to \ell$ energy spectrum provide the strongest 
bounds on such anomalous couplings. The presence of these dipole moments are 
shown to influence, \eg, the extraction of $\alpha_s(m_\tau^2)$ from $\tau$ 
decays and can lead to apparent violations of CVC expectations. 

\end{abstract} 

\vskip0.45in
\begin{center}

Submitted to Physical Review {\bf D}.

\end{center}


\renewcommand{\thefootnote}{\arabic{footnote}} \end{titlepage}


 Although the Standard Model(SM) continues to do an excellent job of 
describing almost all experimental data{\cite {moriond}}, many believe that 
new physics must exist beyond the SM for a number of reasons and that it 
cannot be too far away. In 
addition to searches for new particle production or the observation of 
rare or forbidden decays of well known particles at colliders, an 
alternative probe for new physics is that 
of precision measurements. One possibility is that the repetitive nature of 
the family structure in the SM may provide some insight into what form this 
new physics might take. In particular the detailed properties of the third 
family of SM fermions may be the most sensitive probes of a new mass scale 
which is beyond the direct reach of current accelerators. To this end it is 
important to examine the nature of the $b$ and $t$ quarks as well as the 
$\tau$ lepton with as much precision as possible. 

In the $\tau$ case one has a particularly clean laboratory in the hunt for 
new physics due to the added advantage associated with the fact that 
strong interaction effects 
can only arise at higher order, if at all, depending upon which 
$\tau$ property is being probed. For this reason we focus on the $\tau$ 
lepton in the following discussion. 
One possible form of new physics associated with the $\tau$ is a 
subtle modification of how the $\tau$ interacts with the gauge bosons of the 
SM, \ie, the $Z$, $\gamma$ and $W$. These new interactions, if their scale is 
sufficiently large, can be parameterized as a series of higher dimension, 
gauge invariant--though 
non-renormalizable--operators involving the $\tau$, the SM gauge fields and, 
perhaps, the Higgs boson as well. 
Neglecting the possibility of leptonic CP violation, the new operators of 
lowest dimension that one can construct take 
the form of anomalous magnetic moment-type interactions. To explicitly 
construct these operators using only the SM particle content requires us to 
explicitly introduce the scalar Higgs field{\cite {old}} in order to produce 
the required helicity flip while maintaining gauge invariance. Thus we find 
that anomalous dipole moment operators for the $\tau$ are necessarily of 
dimension-six. 

In the case of the $Z$ and photon, our normalization for these operators take 
the conventional form 
\begin{equation}
{\cal L}_Z={g\over {2c_w}}\bar \tau\left[\gamma_{\mu}
(v_\tau-a_\tau\gamma_5)+{i\over {2m_\tau}}
\sigma_{\mu\nu}q^{\nu}(\kappa_\tau^Z-i\tilde\kappa_\tau^Z\gamma_5)
\right]\tau Z^{\mu} \,,
\end{equation}
where as usual $a_\tau=-1/2$ and $v_\tau=-1/2+2\sin^2 \theta_w$.  
Similarly for the photon we can write an almost identical interaction 
structure: 
\begin{equation}
{\cal L}_\gamma=e\bar \tau\left[Q_\tau\gamma_{\mu}
+{i\over {2m_\tau}}
\sigma_{\mu\nu}q^{\nu}(\kappa_\tau^\gamma-i\tilde\kappa_\tau^\gamma\gamma_5)
\right]\tau A^{\mu} \,,
\end{equation}
where $Q_\tau=-1$. In both cases $\kappa(\tilde \kappa)$ corresponds to an  
anomalous magnetic(electric) dipole complex {\it form factor}. 
Of course as is well-known, radiative corrections in the SM can easily 
induce dipole moment-type interactions. In this specific case, 
since the real part of the $\tilde \kappa$'s are intrinsically CP violating,  
they remain zero to several loops whereas the $\kappa$'s receive complex, order 
$\alpha$, contributions{\cite {sm}}. In both cases the imaginary parts of 
$\kappa$ and $\tilde \kappa$ arise due to the absorptive parts of the loop 
diagrams. In the present paper we will be interested in anomalous 
magnetic dipole type couplings that are over and above those of the SM and 
are comparable or perhaps somewhat larger 
in magnitude. We might expect that if very high mass scales are inducing these 
anomalous couplings then the parameters $\kappa$ and $\tilde \kappa$ arising 
from this new physics will be real and not to be very scale dependent, \ie, 
their values at $q^2=0$ and $M_Z^2$ will be little different. 

Experiments designed to directly probe the couplings $\kappa_\tau^{\gamma,Z}$ 
and $\tilde \kappa_\tau^{\gamma,Z}$ have been performed at LEP and 
elsewhere{\cite {rev}} with only negative results. In fact, in the $Z$ boson 
case we can use the $Z\to \tau^+\tau^-$ decay width, the $\tau$ 
forward-backward asymmetry and the angular distribution of the $\tau$ 
polarization to constrain both $\kappa_\tau^Z$, which we assume to be real, 
and $|\tilde \kappa_\tau^Z|$ through a `radiative corrections' 
analysis{\cite {oldtgr}}. Using the data as presented at 
Moriond 1997{\cite {moriond}}, a modified version of ZFITTER5.0{\cite {bar}} 
and the input values of $\alpha_s(M_Z)=0.118$ and 
$\alpha^{-1}_{EM}=128.896${\cite {alpha}} we obtain the $95\%$ CL regions 
shown in Fig. 1. From this analysis we can conclude that if the anomalous 
couplings are the only source of new physics the $Z$-pole data tells us that 
$|\kappa_\tau^Z,\tilde \kappa_\tau^Z|\leq 0.0023$. Note that this value 
$\sim \alpha/\pi$, the typical size of a SM loop correction. The direct 
search results yield comparable limits, particularly in the case of the 
CP-violating couplings. The corresponding 
bounds on $\kappa_\tau^\gamma$ and 
$\tilde \kappa_\tau^\gamma$ obtained through direct means are somewhat weaker.

\vspace*{-0.5cm}
\nn
\begin{figure}[htbp]
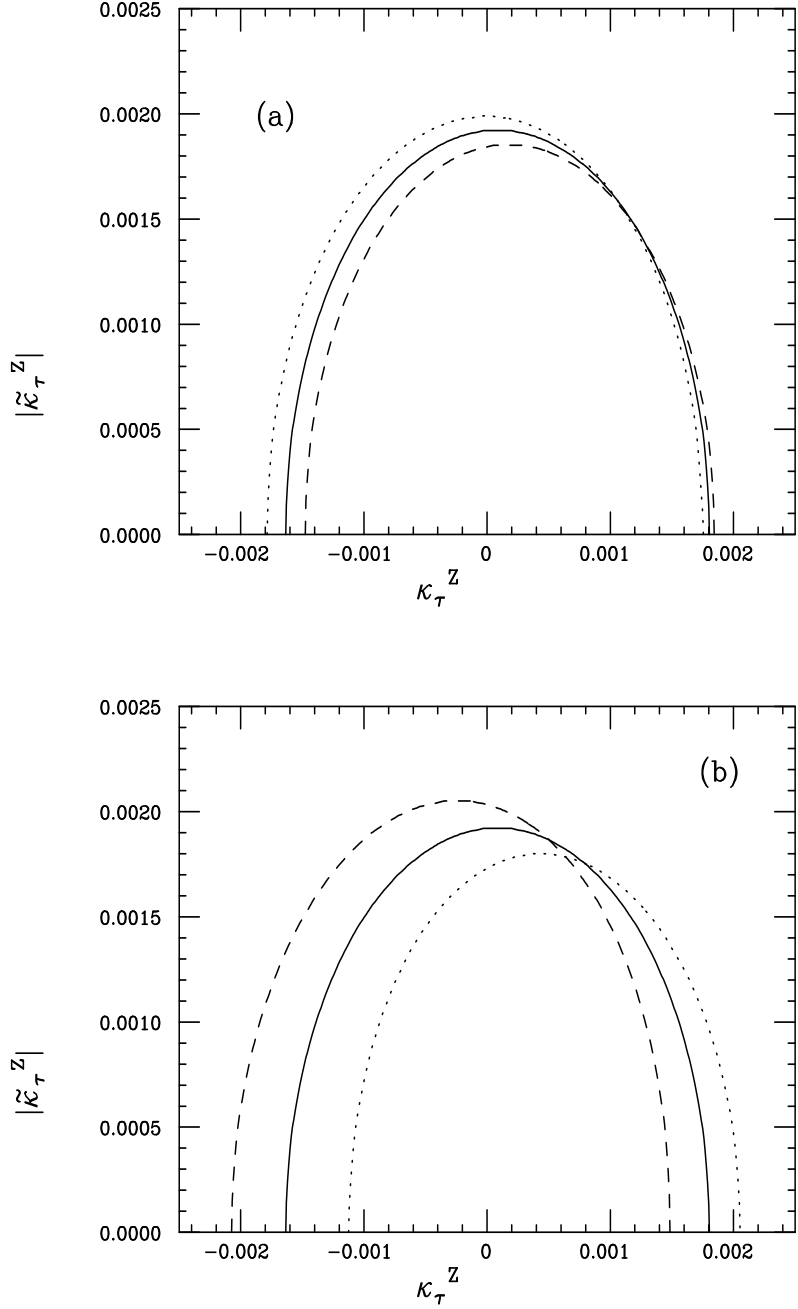

\centerline{
\psfig{figure=chi2tau.fig1ps,height=10cm,width=12cm,angle=-90}}
\vspace*{-.75cm}
\centerline{
\psfig{figure=chi2tau.fig2ps,height=10cm,width=12cm,angle=-90}}
\vspace*{-0.6cm}
\caption{ $95\%$ CL allowed regions for the $\tau \tau Z$ anomalous dipole 
moment couplings. In (a) $m_H=300$ GeV has been assumed and the three curves 
correspond to $m_t$=169(dotted), 175(solid) and 181(dashed) GeV, respectively. 
In (b)  $m_t=175$ GeV is assumed for $m_H$=60(dotted), 300(solid) or 
1000(dashed) GeV, respectively.}
\end{figure}
\vspace*{0.4mm}

The purpose of this paper to examine the analogous CP conserving 
{\it charged current} dipole 
moment interactions of the $\tau$ which have not been as extensively discussed 
in the literature{\cite {cc}}. To this end, 
we modify the charged current $\tau \nu W$ interaction to now be of the form 
\begin{equation}
{\cal L}_{\tau \nu W} = {g\over {\sqrt 2}} \bar \tau \left[\gamma_\mu+
{i\over {2m_\tau}}\kappa_{\tau}^W \sigma_{\mu\nu}q^\nu\right]P_L
\nu_\tau W^\mu +h.c.\,,
\end{equation}
which parallels the above coupling structures for the photon and $Z$. Here 
$P_L$ is the left-handed projection operator and $\kappa_\tau^W$ is assumed 
to be real. If one allows for the possible existence of 
$\kappa_\tau^{Z,\gamma}$ then $\kappa_\tau^W$ must also exist with a 
comparable magnitude due to the demand that the original 
non-renormalizable operators be gauge invariant. This is easily seen in a 
number of ways but perhaps the simplest is to examine a particular realization 
for these non-renormalizable operators as presented in Ref.{\cite {old}}: 
\begin{equation}
{\cal L}_{new} = {1\over {\Lambda^2}} \bar L \sigma_{\mu\nu}
\left[\alpha_{\tau B} B^{\mu\nu}+\alpha_{\tau W}T_aW_a^{\mu\nu}\right]
\tau_R \Phi+h.c.\,,
\end{equation}
Here, $L$ is the left-handed doublet containing the $\tau$, 
$\Lambda$ is a large mass scale, $\alpha_{\tau W, \tau B}$ are a pair of 
parameters, $W_a^{\mu\nu}$ and $B^{\mu\nu}$ are the $SU(2)_L$ and $U(1)$ 
field strength tensors, $T_a$ are the $SU(2)_L$ generators,  and $\Phi$ is 
the Higgs field which we can ultimately replace by 
its vacuum expectation value. Re-writing these operators in the mass 
eigenstate basis and employing the normalizations of Eqs. (1-3) we see that 
we can immediately write a pair of extremely 
simple relations amongst the three anomalous 
magnetic moment type couplings which do not depend upon the scale $\Lambda$:
\begin{eqnarray}
{\kappa_\tau^W\over {\kappa_\tau^Z}}  &=& {-1\over {sc}}{tx\over {t+x}} 
\,, \nonumber \\
{\kappa_\tau^\gamma\over {\kappa_\tau^Z}}  &=& {1\over {2sc}}{tx+1\over {t+x}}
\,, 
\end{eqnarray}
where $t=\sin \theta_w/ \cos \theta_w \equiv s/c$ and 
$x\equiv \alpha_{\tau W}/ \alpha_{\tau B}$. A plot of these ratios is shown 
in Fig. 2 where we see that they are typically of order unity except near the 
points $x=-t$ and $x=0$. These arguments possibly suggest that the relevant 
interesting range of $|\kappa_\tau^W|$ may not be much larger than about 
0.001-0.01 since the corresponding anomalous $\kappa_\tau^Z$ couplings were 
found to 
to be quite small. Again, this range is not much different than what we might 
expect from a typical loop correction of order $\alpha/\pi$ in the SM or 
simple extensions thereof. Since we will concentrate solely on $\kappa_\tau^W$ 
below, we will henceforth employ the simple 
notation $\kappa_\tau^W\equiv \kappa$. 

\vspace*{-0.5cm}
\nn
\begin{figure}[htbp]
\centerline{
\psfig{figure=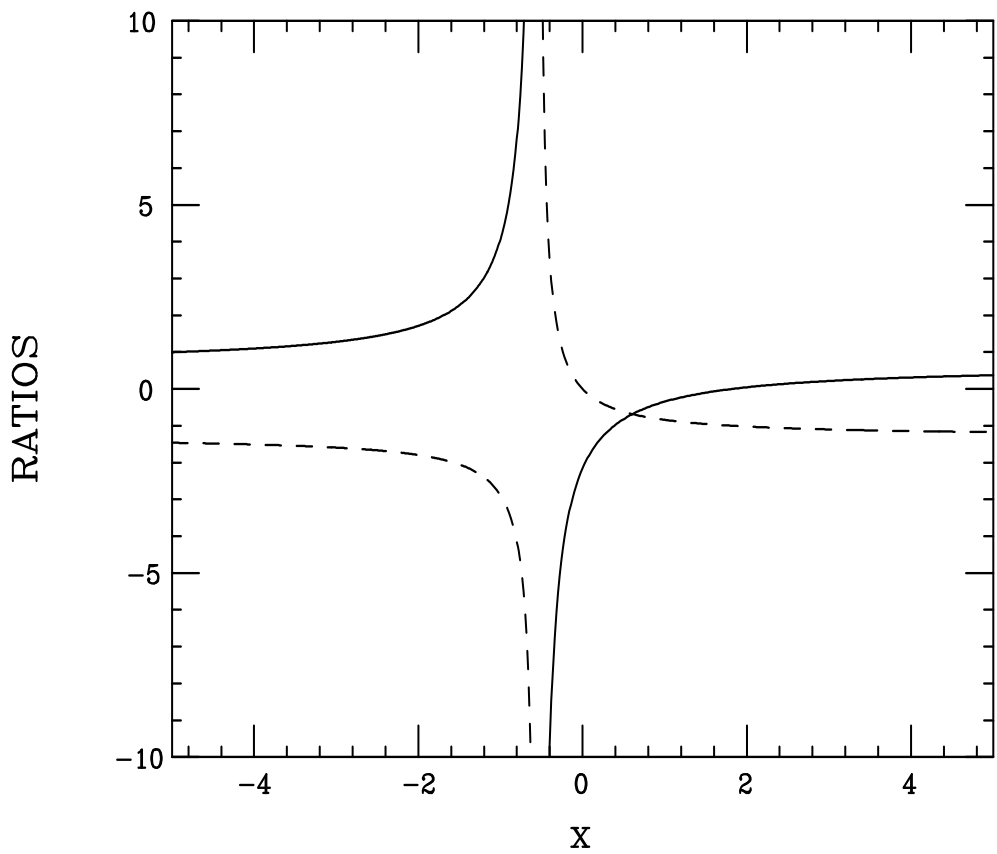,height=14cm,width=16cm,angle=-90}}
\vspace*{-1cm}
\caption{The ratios $\kappa_\tau^W/ \kappa_\tau^Z$ (dash) and 
$\kappa_\tau^\gamma/ \kappa_\tau^Z$ (solid) as functions of the 
parameter $x$. }
\end{figure}
\vspace*{0.4mm}

The first point we need to address is what are the current direct limits 
on $\kappa$ and how far are they away 
from the suggestive range of 0.001-0.01? The most obvious set of constraints 
arise from considerations of lepton universality since the $\tau$ now 
interacts with the $W$ in a away that differs from the other leptons, which we 
treat as having only SM interactions. The two best places to test $\tau$ 
universality are the $\tau$ lifetime itself and the ratio of $W$ decay widths 
$R=\Gamma(W\to \tau \nu)/ \Gamma(W\to \mu \nu ,e \nu)$. It is obvious that 
ratios such as $\Gamma(\tau \to P\nu)/\Gamma(P\to \mu \nu)$, where $P$ is a 
pseudoscalar meson, will not yield additional constraints. Since the matrix 
element of the hadronic weak current between the pseudoscalar and the vacuum 
is proportional to the momentum of the virtual $W$, the 
$\sigma_{\mu\nu}q^\nu$ piece of the $\tau$'s leptonic current will drop out, 
\ie, only the longitudinal part of the virtual $W$ contributes. 
Any deviation from universality observed in such ratios {\it cannot} be 
attributed to an anomalous moment for the $\tau$. 

For the leptonic 
decay of the $\tau$ we find, neglecting the mass of the lepton in the final 
state 
\begin{equation}
\Gamma(\tau \to \ell \nu \bar \nu) = \Gamma_0 (1+{\kappa\over {2}}+
{\kappa^2\over {10}})\,,
\end{equation}
where $\Gamma_0$ is the conventional SM result, while for the $W\to \tau \nu$  
decay width we obtain 
\begin{equation}
\Gamma(W\to \tau \nu) = {G_FM_W^3\over {6{\sqrt 2}\pi}}(1-r)^2\left[1+
{r\over {2}} -{3\kappa \over {4}}+{\kappa^2(1+2r)\over {8r}}\right]\,,
\end{equation}
where $r=(m_\tau/M_W)^2$. Here we see that in the $W$ decay case the 
$\kappa^2$ term is dramatically enhanced kinematically. It is easy to 
include in these equations the effect of a CP violating electric dipole 
moment type term, $\tilde \kappa$, by the replacement of the $\kappa^2$ in the 
last term of both expressions by 
the combination $\kappa^2+\tilde \kappa^2$. $\tilde \kappa$ only appears 
quadratically since neither of these quantities are CP violating observables. 
Using the most recent experimental data{\cite {pich}} we are able to obtain 
the following $95\%$ CL bounds, neglecting potential $\tilde \kappa$ 
contributions:
\begin{eqnarray}
B_\ell \tau_\tau/\tau_\mu  &:& -2.24\cdot 10^{-2}\leq \kappa \leq 
2.31\cdot 10^{-2}\,, \nonumber \\ 
R  &:& -2.54\cdot 10^{-2}\leq \kappa \leq 2.83\cdot 10^{-2}\,, 
\end{eqnarray}
Note that the two constraints yield comparable limits implying that we must 
have $|\kappa|\leq 0.0283 \simeq 0.03$, which is somewhat larger that the 
bounds we obtained 
for the $Z$ case. We chose not to combine these two results since, 
in principle, $\kappa(0)$, which is probed by the $\tau$ width may differ from 
$\kappa(M_W^2)$, probed in $W$ decays. Note that the $W$ decay expressions can 
also be used to obtain a bound on $|\tilde \kappa|$ of a comparable 
size, but any limit obtainable from the $\tau$ lifetime will be rather 
poor. Clearly, we need to improve our sensitivity to nonzero values of 
$\kappa$ by roughly a factor of a few to an order of magnitude. The statistics 
for doing so will be available at future runs of the Tevatron and at future 
$B$ and $\tau/c$ factories. 

Where else might we be able to 
probe values of $\kappa$ of order 0.01 or less? In addition 
to modifying the total $\tau$ leptonic decay width, the presence of 
$\kappa \neq 0$ produces a distortion in the final state lepton spectrum 
which in general {\it cannot} be expressed as shifts in the Michel 
parameters. We recall that the general nature of the Michel spectrum assumes 
the absence of derivative couplings--something we now have due to the 
anomalous moment. 
We find that the invariant double differential decay distribution for $\tau$ 
leptonic decays in  the absence of the polarization dependent terms is given by
\begin{equation}
{d\Gamma\over {dE_\ell d\cos \theta_{\tau \ell}}} \sim 3m_\tau^2(p\cdot k)-
4(p\cdot k)^2(1-{\kappa \over {2}})+{3\over {2}}{\kappa^2\over {m_\tau^2}}
(p\cdot k)(m_\tau^2-2p\cdot k)^2\,,
\end{equation}
where $p(k)$ is the $\tau$'s(final state lepton's) four momentum and terms of 
order $m_\ell^2/m_\tau^2$ have been neglected. Note that 
in the $\kappa^2$ term cubic powers of $p\cdot k$ appear which are clearly 
incompatible with the standard Michel spectrum. As above, we can include the 
CP violating electric dipole contribution by the standard replacement 
$\kappa^2 \to \kappa^2+\tilde \kappa^2$, which implies that the lepton 
energy spectrum is not very sensitive to $\tilde \kappa$ non-zero as we 
might expect. There are two places where precision 
measurements of the $\tau$ lepton spectrum can be made: at high energies,  
sitting on the $Z$, or at low energies at a $B$ or $\tau/c$ factory. On the 
$Z$ an extra advantage is obtained due to the fact that the $\tau$ is 
naturally polarized. Beam polarization can greatly enhance this added 
sensitivity as has been exploited by the SLD Collaboration{\cite {sld}}. 
At the $Z$, the 
{\it normalized} lepton energy spectrum for the $\tau$ to leading order in 
$\kappa$, which seems a reasonable approximation since $\kappa$ is small, 
can be written in the absence of $m_\tau^2/M_Z^2$ and $m_\ell^2/m_\tau^2$ 
corrections as
\begin{equation}
{1\over {N}} {dN\over {dz}} = f(z)+P^\tau_{eff}g(z)\,, 
\end{equation}
where $z=E_\ell/E_\tau$, $P^\tau_{eff}$ is the production 
angular-dependent effective polarization of the 
$\tau$ including the effects of the initial $e^-$ beam polarization as 
given in Ref.{\cite {sld}} and $f,g$ are kinematic functions:
\begin{eqnarray}
f  &=& ({5\over {3}}-3z^2+{4\over {3}}z^3)+\kappa(-{1\over {6}}+{3\over {2}}
z^2-{4\over {3}}z^3)\,, \nonumber \\
g  &=& (-{1\over {3}}+3z^2-{8\over {3}}z^3)+\kappa(-{1\over {2}}+2z
-{3\over {2}}z^2)\,, 
\end{eqnarray}
The behaviour of both $f(z)$ and $g(z)$ due to variations in $\kappa$ is 
shown in Fig. 3. There are several things to observe: first, $g$ is more 
sensitive to variations in $\kappa$ than is $f$ so that having a handle on 
$P^\tau_{eff}$ through modifications of the beam polarization is very 
useful in obtaining a greater sensitivity. Second, the $\kappa$ dependent term 
in $f$ has the same structure as that due to a shift in 
${8\over {3}}\cdot \rho$, where $\rho$ is one of the conventional Michel 
parameters{\cite {ajw}}. The $\kappa$ dependent term in $g$ is {\it not} of 
the same form as a shift in the corresponding $\delta$ parameter. 

\vspace*{-0.5cm}
\nn
\begin{figure}[htbp]
\centerline{
\psfig{figure=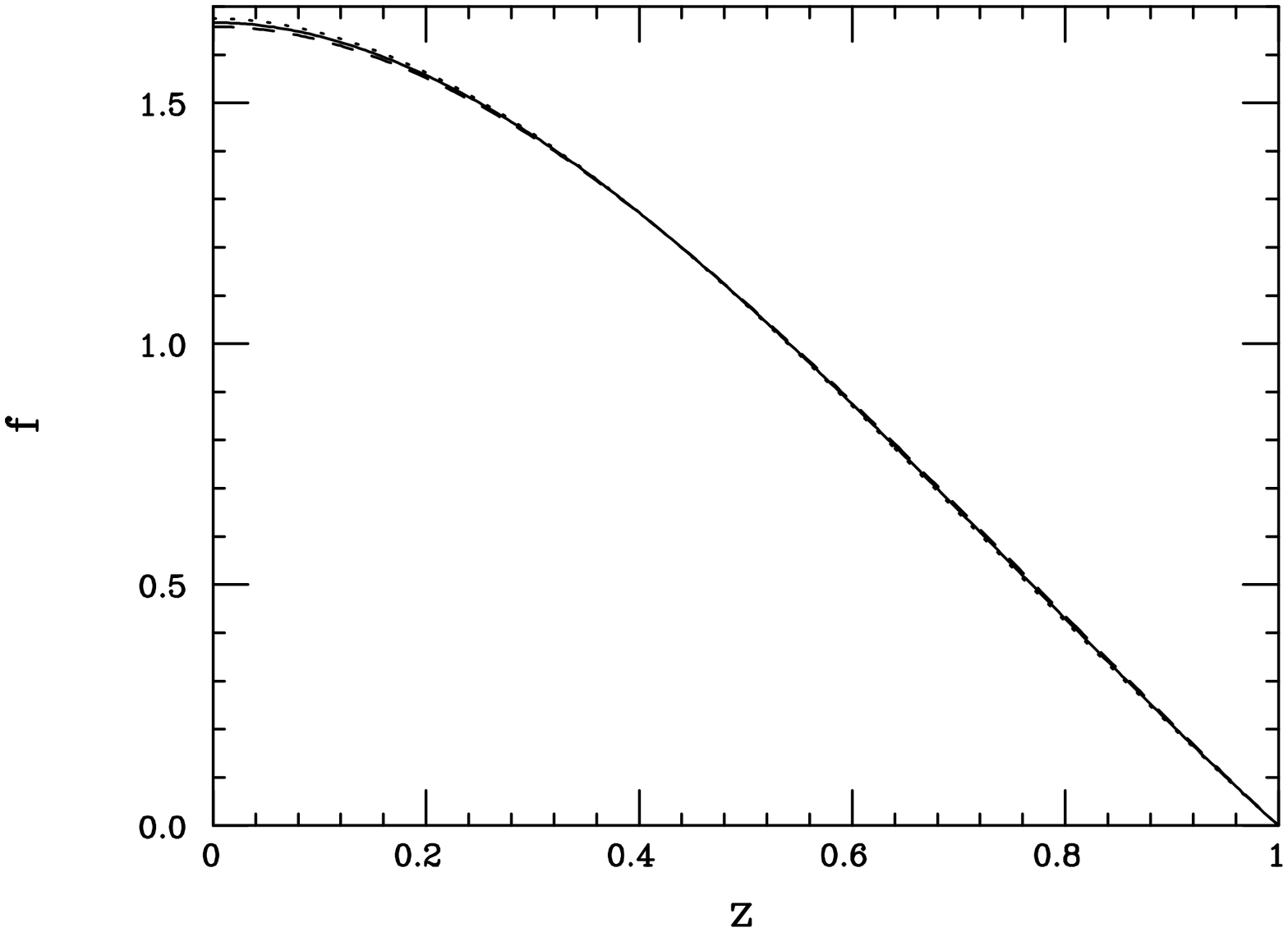,height=10cm,width=12cm,angle=-90}}
\vspace*{-.75cm}
\centerline{
\psfig{figure=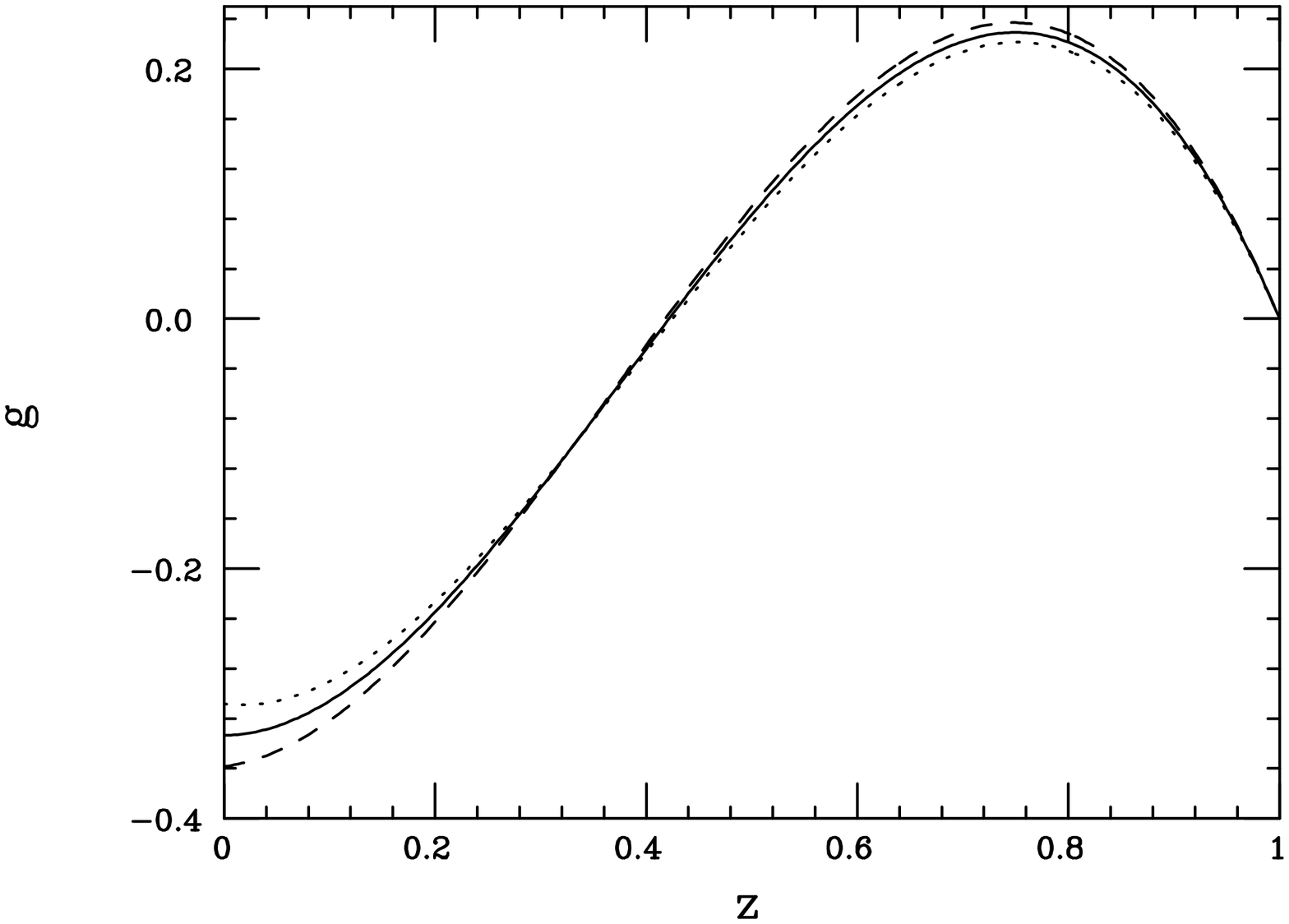,height=10cm,width=12cm,angle=-90}}
\vspace*{-0.6cm}
\caption{The kinematic functions $f(z)$ and $g(z)$ for $\kappa$ values of 
-0.05(dot), 0(solid), or 0.05(dash).}
\end{figure}
\vspace*{0.4mm}

The current bounds on $\kappa$ from the SLD Collaboration 
are rather poor due to the low 
statistics available. To get an idea of what future constraints on $\kappa$ 
might be obtainable, we perform a toy Monte Carlo study assuming a sample of 
20,000 $\tau \to \ell$ decays, corresponding to about 1.5 million $Z$'s,  
assuming a beam polarization of $|P|=77.2\%$ and a cut 
on the angular acceptance of $|\cos \theta_\ell| \leq 0.75$. We follow the 
same approach as SLD{\cite {sld}} and consider four distinct cases depending 
on whether $P$ is positive or negative and whether the negatively charged 
$\tau$ goes into the forward or backward hemisphere of the detector. 
In each case, we divide 
the $z$ range into 10 equal bins, generating data weighted by the statistical 
errors only. Fig. 4 shows the resulting Monte Carlo generated distributions in 
comparison with the expectations of the SM. 

\vspace*{-0.5cm}
\nn
\begin{figure}[htbp]
\centerline{
\psfig{figure=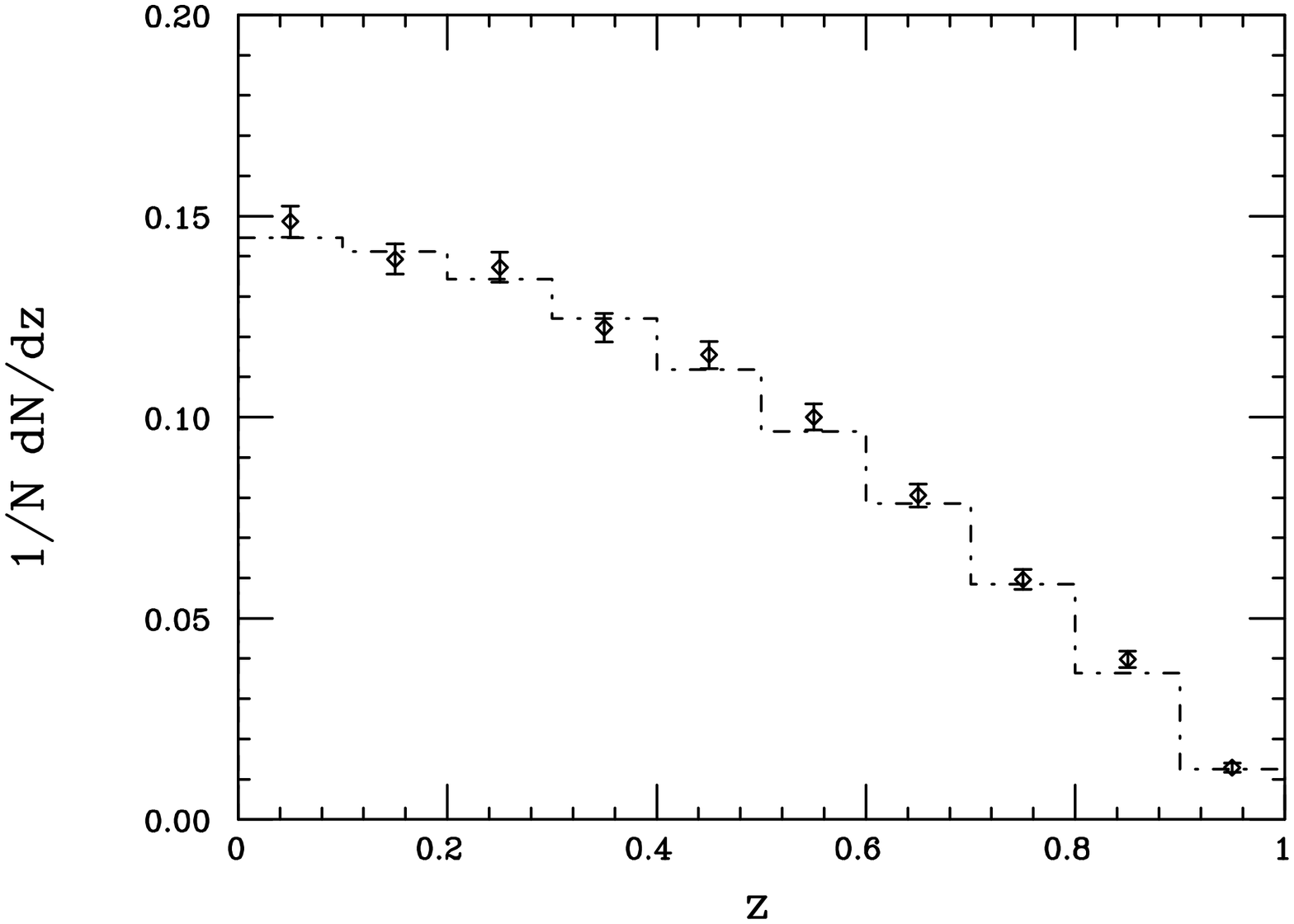,height=10cm,width=12cm,angle=-90}}
\vspace*{-.75cm}
\centerline{
\psfig{figure=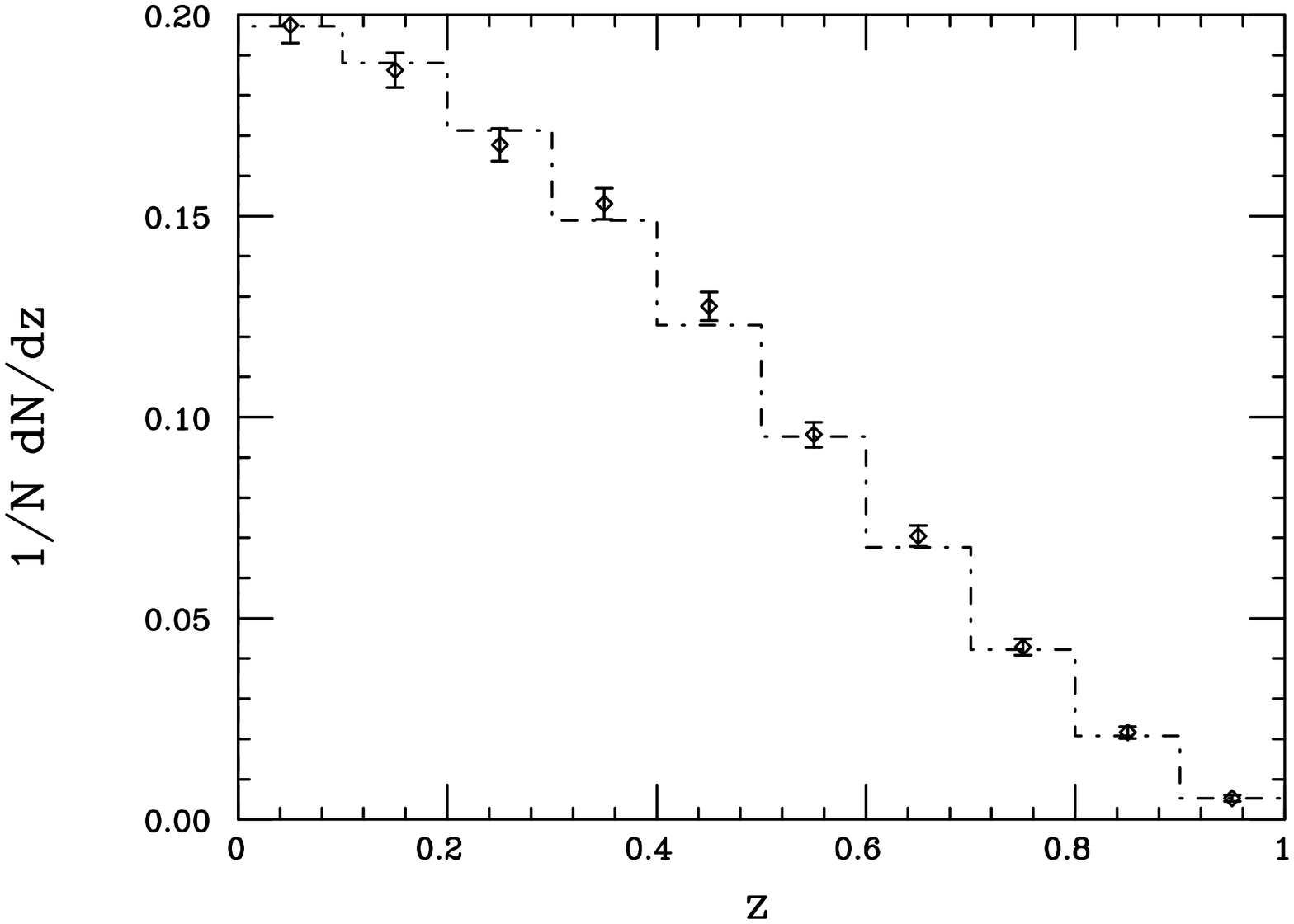,height=10cm,width=12cm,angle=-90}}
\vspace*{-0.6cm}
\caption{Comparison of SM expectations and Monte Carlo data for the lepton 
energy spectrum in $\tau$ decays at the $Z$. The top figure 
corresponds to the sum of the 2 cases where the negative $\tau$ is forward and 
$P>0$ and where the $\tau$ is backward and $P<0$. The bottom plot shows the 
sum of the other two possibilities. 
In both figures the SM expectations are given by the dash-dotted 
histogram whereas the points represent the Monte Carlo data generated assuming 
the SM is correct.}
\end{figure}
\vspace*{0.4mm}

In order to obtain our $\kappa$ constraint, we perform a $\chi^2$ fit to 
the above distributions, which were generated assuming $\kappa=0$, by 
using the $\kappa$-dependent functional forms above. We obtain a best fit of 
$\kappa=0.06\pm 0.12$ at $95\%$ CL with a $\chi^2/d.o.f.$ of 41.8/40; this 
result is explicitly shown in Fig.5. It is clear from this analysis that $Z$ 
pole measurements will never be able to achieve the level of sensitivity we 
require to probe $|\kappa|$ values below 0.03.

\vspace*{-0.5cm}
\nn
\begin{figure}[htbp]
\centerline{
\psfig{figure=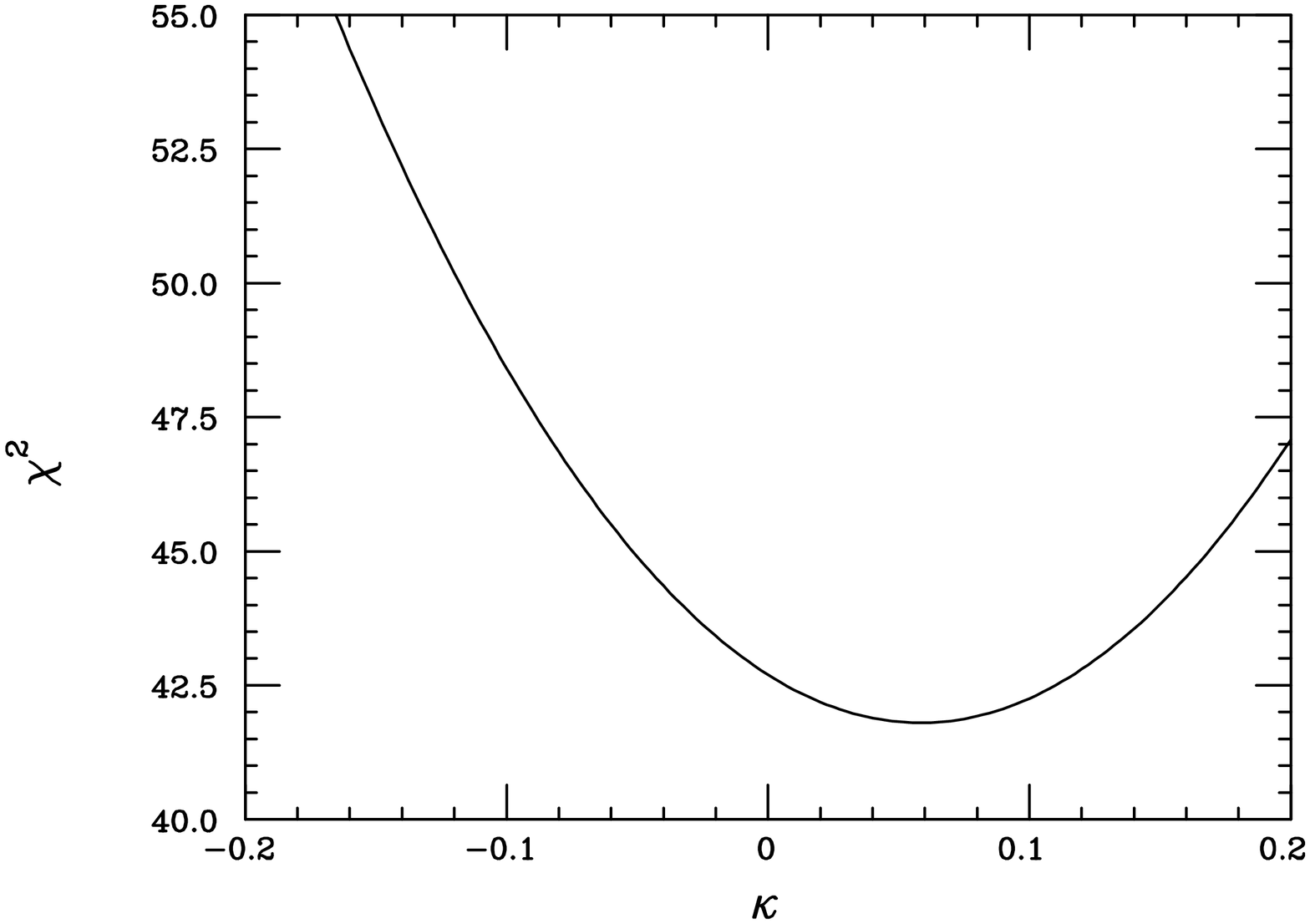,height=14cm,width=16cm,angle=-90}}
\vspace*{-1cm}
\caption{$\chi^2$ fit to the Monte Carlo data using the $\kappa$-dependent 
form of the leptonic energy spectrum in $\tau$ decays on the $Z$ pole.}
\end{figure}
\vspace*{0.4mm}

What happens at a $B$ or $\tau-c$ factory? Here the $\tau$ is either at rest or 
is relatively slow and a pseudo-rest frame can be defined. 
In the $\tau$ rest frame, neglecting the $\ell$ mass and polarization dependent 
terms, one finds
\begin{equation}
{d\Gamma(\tau \to \ell \nu \bar \nu)\over {dx}} = \Gamma_0 x^2\left[6-4x
+2\kappa x+3\kappa^2(1-x)^2\right]\,,
\end{equation}
where $x=2E_\ell/m_\tau$ and $\Gamma_0$ was introduced above. (As above, the 
contribution of $\tilde \kappa$ can be included by the replacement 
$\kappa^2 \to \kappa^2+\tilde \kappa^2$.) Neglecting terms 
of order $\kappa^2$ again we see that the 
{\it normalized} lepton energy spectrum in the $\tau$ rest frame can be 
written as
\begin{equation}
{1\over {\Gamma}} {d\Gamma\over {dx}} = x^2\left[6-4x+\kappa (4x-3) \right]\,,
\end{equation}
As in the $Z$ case, we see that by 
comparing with the conventional Michel spectrum{\cite {pich}} the 
effect of $\kappa$ on the normalized spectrum {\it to this order} is the same 
as $8/3 \delta \rho${\cite {ajw}}. The current best single measurement of 
$\rho$ using the final state lepton spectrum comes from CLEO{\cite {cleo}}; 
assuming 
$e-\mu$ universality they obtain $\delta \rho=\rho-0.750=-0.015\pm 0.015$ or 
$\kappa =-0.040\pm 0.040$. This is already far better than what we obtained 
earlier on the $Z$ peak using our toy Monte Carlo. Using the current 
world average value 
$\delta \rho=\rho-0.750=-0.009\pm 0.014${\cite {evans}} gives a comparable 
result. It is clear that future $B$-factories, with more than 
an order of magnitude increase in statistics should begin to probe $\kappa$ 
values of order 0.01. Fig. 6 shows a plot of the function 
$h(x)={1\over {\Gamma}} {d\Gamma\over {dx}}$ for different values of $\kappa$. 
Note that the most significant deviation from the SM occurs at large values 
$x$ where the most statistics are available.

\vspace*{-0.5cm}
\nn
\begin{figure}[htbp]
\centerline{
\psfig{figure=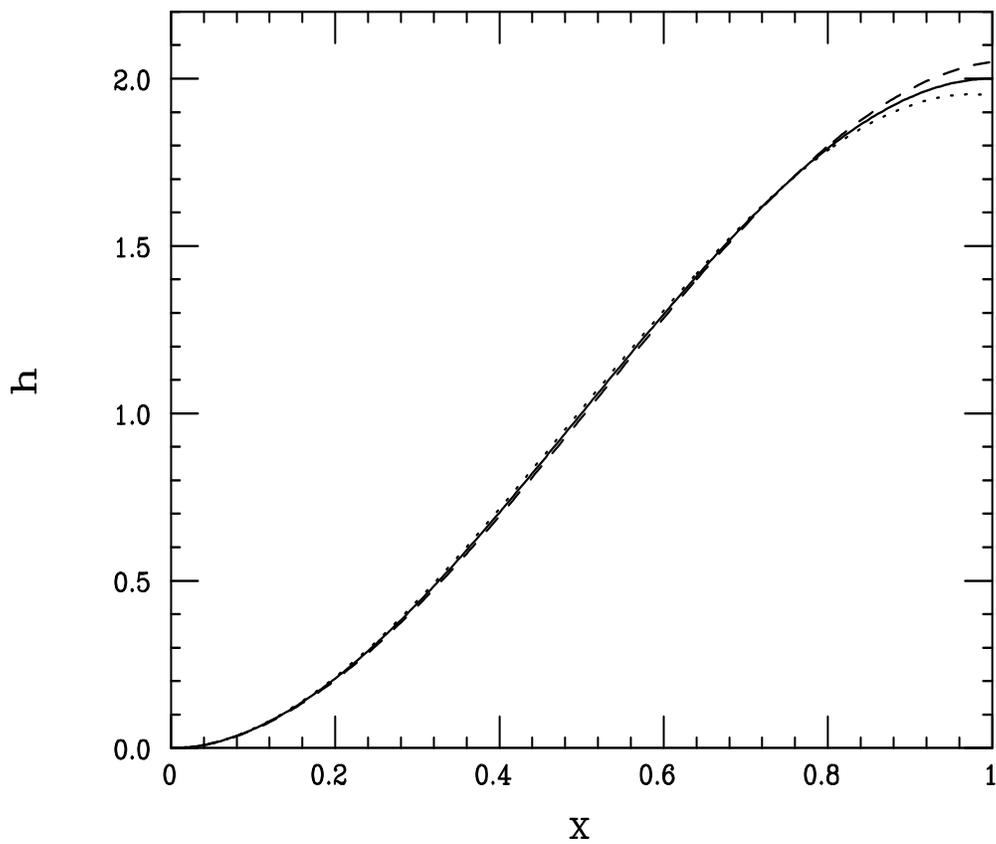,height=14cm,width=16cm,angle=-90}}
\vspace*{-1cm}
\caption{h(x) as defined in the text for $\kappa$=-0.05(dotted), 0(solid), 
and 0.05(dashed).}
\end{figure}
\vspace*{0.4mm}

In order to get an estimate of the future sensitivity of low energy 
measurements to non-zero values of $\kappa$, we return to our Monte Carlo 
approach assuming a sample of $10^8$ $\tau$ pairs. Data on the $\tau$'s lepton 
energy spectrum is then generated assuming the SM is correct and is put into 
10 equal sized $x$ bins as is shown in Fig. 7. Only statistical errors were 
included.

\vspace*{-0.5cm}
\nn
\begin{figure}[htbp]
\centerline{
\psfig{figure=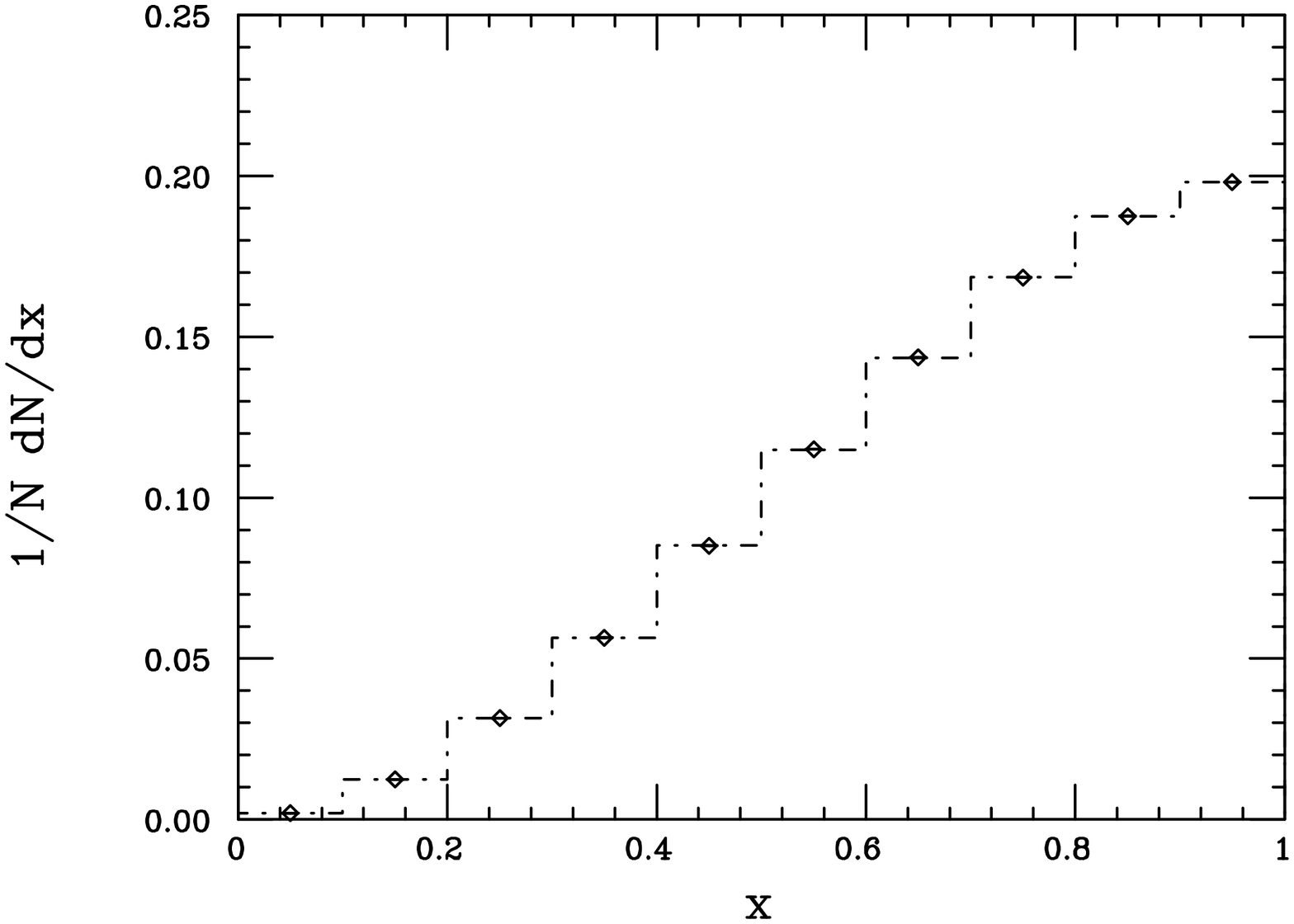,height=14cm,width=16cm,angle=-90}}
\vspace*{-1cm}
\caption{Comparison of SM expectations and Monte Carlo data for the lepton 
energy spectrum in $\tau$ decays at a $B$ or $\tau-c$ factory.  
As above, the SM expectations are given by the dash-dotted 
histogram whereas the points represent the Monte Carlo data generated assuming 
the SM is correct.}
\end{figure}
\vspace*{0.4mm}

A $\kappa$-dependent fit is the performed with the results shown in Fig. 8. 
Here we see that the fitted value of $\kappa$ is determined to be 
$\kappa=(-0.49^{+0.82}_{-0.81})\cdot 10^{-3}$ at $95\%$ CL with a 
$\chi^2/d.o.f.$ of 11.9/10. A parallel analysis using a data sample of the 
same size but with ten times as many bins obtained 
$\kappa=(-0.04\pm 0.79)\cdot 10^{-3}$ at $95\%$ CL with a $\chi^2/d.o.f.$ of 
90.5/100. From this analysis it is clear that very large $\tau$ data samples 
of this magnitude will allow the probing $\kappa$'s in the 0.001-0.01 range 
even if significant systematic errors are present. 

\vspace*{-0.5cm}
\nn
\begin{figure}[htbp]
\centerline{
\psfig{figure=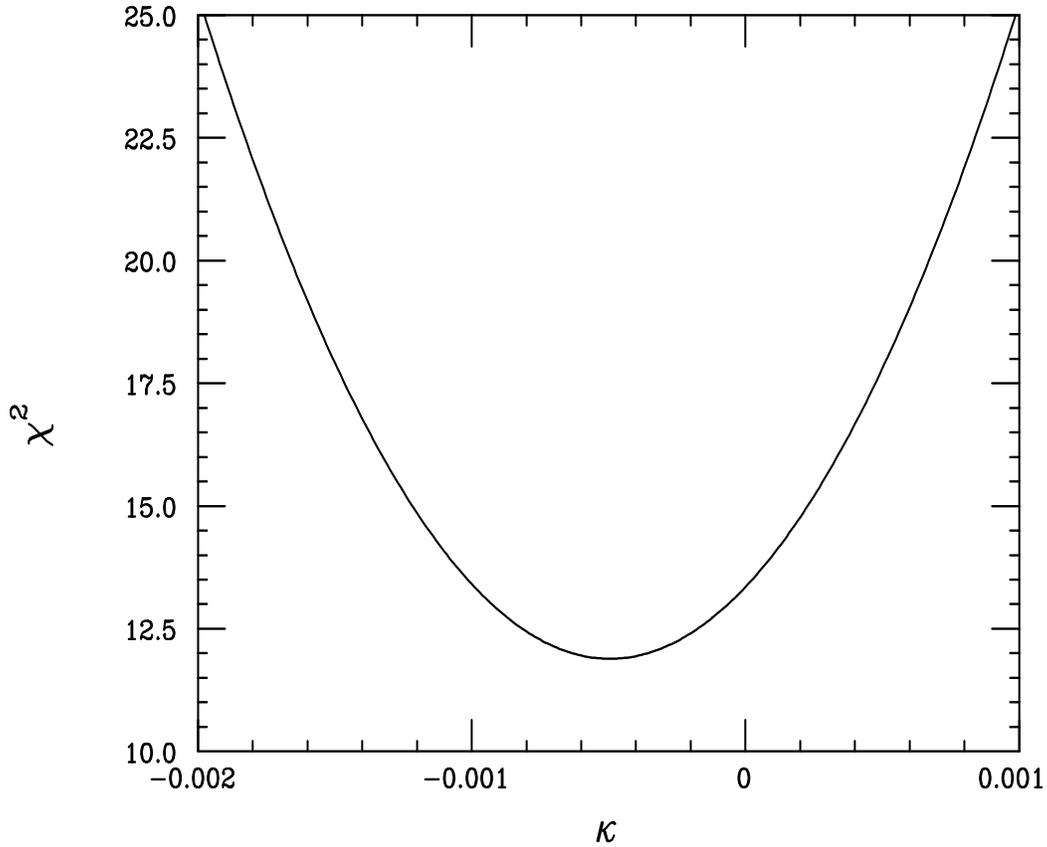,height=14cm,width=16cm,angle=-90}}
\vspace*{-1cm}
\caption{$\chi^2$ fit to the Monte Carlo data using the $\kappa$-dependent 
form of the leptonic energy spectrum in $\tau$ decays at a $B$ or $\tau-c$ 
factory.}
\end{figure}
\vspace*{0.4mm}

Are there additional constraints that we can obtain on $\kappa$ from the 
semileptonic decay modes? As we saw above, $\tau$ decays to pseudoscalar 
mesons do not provide any additional sensitivity to $\kappa \neq 0$ 
since the purely transverse $\sigma_{\mu\nu}q^\nu$ terms 
decouple in this case. We thus turn to decays of the type $\tau \to V\nu$ 
where $V$ is either a vector or axial-vector meson or where $V$ 
represents the hadronic continuum at and above the two pion threshold. As is 
well known, the ratio of the semileptonic to leptonic widths of the $\tau$ 
can be symbolically written as 
\begin{equation}
R_\tau={\Gamma(\tau \to hadrons)\over {\Gamma(\tau \to \ell \nu \bar \nu)}} 
\sim \int_0^{m_\tau^2} R({s\over {m_\tau^2}}){\rho(s)ds\over {m_\tau^2}}\,,
\end{equation}
where $\rho(s)=Im\Pi(s)/\pi$ is the spectral density defined in terms of the 
appropriate charged current correlator $\Pi(q^2)$ and $R$ is the 
kinematic kernel which 
now depends on $\kappa$. Defining $x=s/m_\tau^2$ we find that $R(x)$ is given 
to all orders in $\kappa$ by
\begin{equation}
R(x)=(1-x)^2{\left[1+2x-3\kappa x+{1\over {4}}\kappa^2 x(2+x)\right]\over 
{\left[1+{1\over {2}}\kappa+{1\over {10}}\kappa^2\right]}}\,,
\end{equation}
As usual, we can incorporate $\tilde \kappa$ contributions by the now standard 
replacement $\kappa^2 \to \kappa^2+\tilde \kappa^2$ in this expression; again, 
the numerical contributions from the  
$\tilde \kappa$ terms cannot be large. The above expression for $R(x)$, 
expanded to linear order in $\kappa$ and neglecting possible 
$\tilde \kappa$ contributions is shown in 
Fig. 9 and displays reasonable sensitivity to $\kappa$. In principle, a 
non-zero value of $\kappa$ can be responsible for an apparent breakdown in 
the usual CVC predictions in that the $\tau$'s weak coupling to the $W$ is no 
longer solely a mixture of conventional vector and axial-vector currents. 
Of course, the matrix element of the weak hadronic charged current is still 
simply related to the corresponding electromagnetic 
current by an isospin rotation. However, when it is contracted with the 
$\tau$'s charged current matrix element, new $\kappa$-dependent terms arise 
which are not present in $e^+e^- \to hadrons$. It is clear from Fig. 9 that 
positive(negative) values of $\kappa$ will produce a decrease(increase) in the 
predicted value of the relative hadronic branching fraction of the $\tau$, 
$R_\tau$. As an estimate of the current sensitivity we compare the 
experimental rate for $\tau \to 2\pi \nu$ with that expected from CVC using 
the low-energy $e^+e^-$ data as input{\cite {cvc}}in the region of the $\rho$ 
resonance allowing for $\kappa$ to be non-zero; we find that 
$\kappa=-0.045\pm 0.034$ at $1\sigma$. Also, crudely, we find that a value of 
$\kappa=0.03(-0.03)$ leads to {\it decrease(increase)} in the continuum $\tau$ 
hadronic width of roughly $-3(3)\%$, a value not too much smaller than the 
leading inclusive QCD correction to the tree-level SM electroweak result 
of $\alpha_s(m_\tau^2)/\pi$. $\kappa \neq 0$ could thus have important 
implications to the problems associated with a clear understanding of the 
running of the QCD coupling{\cite {phil}} if the value extracted from 
hadronic $\tau$ decays{\cite {qcd}} is indeed shifted due to these anomalous 
couplings.

\vspace*{-0.5cm}
\nn
\begin{figure}[htbp]
\centerline{
\psfig{figure=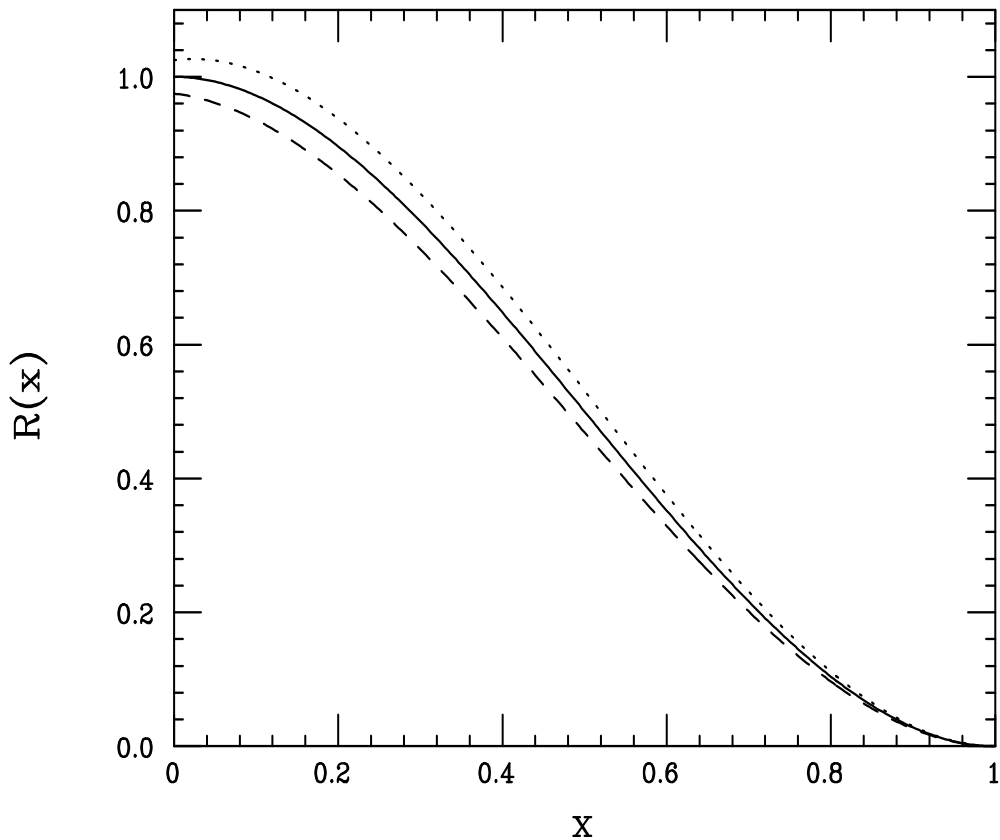,height=14cm,width=16cm,angle=-90}}
\vspace*{-1cm}
\caption{The kinematic function $R(x)$ as defined in the text assuming 
$\kappa$=-0.05(dotted), 0(solid), and 0.05(dashed).}
\end{figure}
\vspace*{0.4mm}

Lastly, we note that a non-zero value for $\kappa$ also leads to a 
modification of the transverse and longitudinal polarization 
fractions for the `$V$' in the final state of this semileptonic decay. 
Unfortunately, the $\kappa \neq 0$ contribution is quite small 
particularly in the low invariant mass region. With 
$x$ now denoting the ratio $m_V^2/m_\tau^2$ we find that to lowest order in 
$\kappa$  
\begin{eqnarray}
F_T  &=& {(2-3\kappa)x\over {1+(2-3\kappa)x}}\,, \nonumber \\
F_L  &=& {1\over {1+(2-3\kappa)x}}\,, 
\end{eqnarray}
where $F_{T(L)}$ is the transverse(longitudinal) fraction of `$V$'. For 
$|\kappa| \leq 0.03$ we see that this effect is below the $1\%$ level; there 
is little sensitivity here to anomalous couplings.

In this paper we have considered how existing and future measurements can 
probe the CP conserving anomalous magnetic dipole moment of the $\tau$, 
$\kappa$, in charged current interactions. Searches for the corresponding 
coupling structure in electromagnetic and weak neutral current interactions 
of $\tau$'s have been undertaken for some time and interesting limits have 
been obtained. We have argued that if $\gamma$ and/or $Z$ anomalous couplings 
are present for the $\tau$ at some level then the analogous charged current 
couplings must also be present with a comparable magnitude due to gauge 
invariance. In almost all cases, except for the decay $W\to \tau \nu$, 
the contributions to the observables that we consider 
from potential CP violating terms arising from the 
electric dipole moment $\tilde \kappa$, if they are at all present, 
are shown to be small. 

We observed that the current best limits on $\kappa$ are indirect and arise 
from universality tests, specifically, considerations of the $\tau$ lifetime 
and the $W\to \tau \nu$ branching fraction. These two independent measurements 
tell us that $|\kappa|\leq 0.028$ at $95\%$ CL. Clearly, at both $B$ and 
$\tau-c$ factories, universality tests involving the $\tau$ should improve 
these indirect limits by factors of order a few to as much as an order 
of magnitude depending on the size of 
systematic errors. More interestingly, direct experiment sensitivity to 
non-zero values of $\kappa$ arise from a number of sources, in particular, the 
shape of final state lepton energy spectrum in $\tau \to \ell \nu \bar \nu$. 
We showed that the shift in the {\it unnormalized} spectrum cannot be 
accommodated through variations in the Michel parameters. However, to first 
order in $\kappa$, the normalized spectrum modification was found to be 
proportional to $\delta \rho$. A toy Monte Carlo study showed that the 
sensitivity obtainable from future data on a sample of more than $10^6$ 
polarized $Z$'s at SLD cannot probe $\kappa$ values at the requisite 0.01 
level. A measurement of the same spectrum at $B$ and/or $\tau-c$ factories 
where $10^8$ $\tau$ pairs are available was shown by the same toy Monte Carlo 
approach to be sensitive to $\kappa$ at the 0.001 level. 

Hadronic $\tau$ decays were shown to display somewhat more subtle sensitivity 
to finite 
$\kappa$ and, in particular, decays to a single pseudoscalar were shown to 
have no sensitivity whatsoever since only the couplings of the virtual 
longitudinal $W$ are being probed there. Decays to `vector' or `axial-vector' 
meson final states(\ie, the hadronic continuum above the $2\pi$ threshold) 
were shown to display three-fold $\kappa$ sensitivity via modifications in 
the overall decay rates as well as in the associated invariant mass 
distributions and the polarization of the final state. In particular we showed 
that the deviations in the value of $R_\tau$ due to a non-zero $\kappa$ can 
lead to an incorrect extraction of the strong coupling constant, 
$\alpha_s(m_\tau^2)$, with the obvious implications elsewhere. We also found 
that the anomalous moment terms can lead to potentially large apparent 
violations of CVC.

\vskip.25in
\begin{center}
{\bf Acknowledgements}
\end{center}
\noindent

The author would like to thank A. Weinstein, CLEO Collaboration, and 
J.L. Hewett for discussions related to this work.

\newpage

%
\def\MPL #1 #2 #3 {Mod.~Phys.~Lett.~{\bf#1},\ #2 (#3)}
\def\NPB #1 #2 #3 {Nucl.~Phys.~{\bf#1},\ #2 (#3)}
\def\PLB #1 #2 #3 {Phys.~Lett.~{\bf#1},\ #2 (#3)}
\def\PR #1 #2 #3 {Phys.~Rep.~{\bf#1},\ #2 (#3)}
\def\PRD #1 #2 #3 {Phys.~Rev.~{\bf#1},\ #2 (#3)}
\def\PRL #1 #2 #3 {Phys.~Rev.~Lett.~{\bf#1},\ #2 (#3)}
\def\RMP #1 #2 #3 {Rev.~Mod.~Phys.~{\bf#1},\ #2 (#3)}
\def\ZP #1 #2 #3 {Z.~Phys.~{\bf#1},\ #2 (#3)}
\def\IJMP #1 #2 #3 {Int.~J.~Mod.~Phys.~{\bf#1},\ #2 (#3)}

\end{document}